\documentclass[preprint, amsmath,amssymb,nofootinbib]{revtex4}
\usepackage{graphicx}
\begin{document}
\bibliographystyle{unsrt}

\title{Creating Statistically Anisotropic and Inhomogeneous Perturbations}

\author{C. Armendariz-Picon}
\email{armen@phy.syr.edu}
\affiliation{Physics Department, Syracuse University,\\ Syracuse, NY13244-1130, USA.}

\begin{abstract}
In almost all structure formation models, primordial perturbations are created within a homogeneous and isotropic universe, like the one we observe. Because their ensemble averages inherit the symmetries of the spacetime in which they are seeded,  cosmological perturbations then happen  to be statistically isotropic and homogeneous. Certain anomalies  in the cosmic microwave background on the other hand suggest that perturbations do not satisfy these statistical properties, thereby challenging perhaps  our understanding of  structure formation.  In this article we relax this tension.  We show that if the universe contains an appropriate triad of scalar fields with spatially constant but non-zero gradients,  it is possible to generate statistically anisotropic and inhomogeneous primordial perturbations, even though the energy momentum tensor of the triad itself is invariant under translations and rotations. 
\end{abstract}

\maketitle

\section{Introduction}
The cosmological principle states that the universe is homogeneous and isotropic on large scales.  Although  it was originally invoked for the sake of simplicity,  it is by now solidly  grounded by observations. A Friedman-Robertson-Walker metric  accurately describes the evolution of the universe on large scales, and it suffices to explain phenomena that range from the abundance of the light elements to the (near) isotropy of the cosmic microwave background. However, homogeneity and isotropy are clearly just approximations. Because of the same cosmological principle, we are able to describe inhomogeneities and anisotropies in our universe as small perturbations around the lowest order homogeneous and isotropic background. 

There is a generalized sense of homogeneity and isotropy that may also apply to these perturbations, namely, \emph{statistical} isotropy and homogeneity.  We shall not extensively dwell on these concepts, and simply note that perturbations are statistically homogeneous and isotropic if their ensemble averages are invariant under translations and rotations (see for instance \cite{Armendariz-Picon:2005jh}.) Given that the cosmological principle appears  to be solidly grounded by observations at this point, it is time to inquire---instead of postulating---whether perturbations also satisfy the generalized, statistical, cosmological principle.   Certain anomalies in the temperature anisotropies in the CMB actually suggest that the answer is negative \cite{deOliveira-Costa:2003pu,Eriksen:2003db,Schwarz:2004gk,Land:2005ad}.

If perturbations and the underlying background were completely independent from each other, the symmetries of the latter would not have to  bear any relation with the statistical symmetries of the former.  However, this independence is not in agreement with our ideas about the origin of structure. According to our present understanding, the structures that we observe  originated from tiny primordial perturbations that were generated during a seeding stage in the early universe \cite{inflation, alternatives}. As a result, primordial perturbations always seem to carry the symmetries of the background on top of which they were seeded. Inflation for instance is typically driven by a homogeneous scalar field, and as consequence,  primordial perturbations created during inflation are statistically homogeneous and isotropic. It is also possible to seed primordial perturbations in an anisotropic but homogeneous universe, and, in agreement with our claim, the resulting primordial perturbations turn out to be  statistically homogeneous but anisotropic \cite{Gumrukcuoglu:2006xj,Ackerman:2007nb}. If the universe is  inhomogeneous during the seeding stage,  the statistical properties of primordial perturbations will accordingly reflect those spatial variations \cite{Donoghue:2007ze}.  

The conclusion that seems to emerge from our previous discussion is that if the universe is homogeneous and isotropic during the seeding state, primordial perturbations have to be statistically homogeneous and isotropic. In this article we argue that this conclusion does not necessarily hold. We present a class of models for the origin of structure  in which the universe is exactly homogeneous and isotropic throughout the seeding stage, with the peculiarity that the resulting primordial perturbations are statistically  anisotropic or inhomogeneous.

\section{A scalar triad}
The class of models that we are going to consider bears a significant relation to previous works in the literature. It contains  a set of three  arbitrary  non-canonical scalar fields $\varphi_\alpha$ (where $\alpha=1,2,3$) minimally coupled to Einstein gravity \cite{Fischler:2004ti,Armendariz-Picon:1999rj}. The action of this ``scalar triad"  is 
\begin{equation}\label{eq:action}
	S=\int d^4 x \sqrt{-g} \left[-\frac{M_P^2}{2} R+\sum_{\alpha=1}^3 L_\alpha(X_\alpha)+\mathcal{L}_m\right], 
\end{equation}
where the $L_\alpha$ are functions of a single argument\footnote{To simplify comparisons with the literature  we work with a $(+,-,-,-)$ metric signature.}
\begin{equation}\label{eq:X}
	X_\alpha=\frac{1}{2}g^{\mu\nu} \partial_\mu \varphi_\alpha \, \partial_\nu \varphi_\alpha,
\end{equation}
and $\mathcal{L}_m$ is the Lagrangian of the remaining matter fields $\psi$, which can include couplings to the triad.  From a different perspective, the triad can be regarded as part of the gravitational sector; the action (\ref{eq:action}) then leads to  what one could interpret as a modification of gravity \cite{Dubovsky:2004sg}.    Note that throughout this work  we use Einstein's summation convention; a sum over indices is implied only if they are in opposite positions.  Otherwise, the location of internal and spatial indices does not have any significance.  

The stability of the system described by the action (\ref{eq:action}) imposes conditions on the functions $L_\alpha$. The requirement that perturbations around any background value of $X_\alpha$ have a positive definite Hamiltonian (in order for the theory to be ghost-free) implies \cite{Aharonov:1969vu}
\begin{equation}\label{eq:stability}
	\frac{dL_\alpha}{dX_\alpha}>0 \quad \text{and} \quad \frac{dL_\alpha}{dX_\alpha}+2 X_\alpha \frac{d^2L_\alpha}{dX^2_\alpha}>0.
\end{equation}
In turn, the last two conditions guarantee that  the system has a well-defined initial value formulation, at least locally, that is, that scalar perturbations  propagate around appropriate characteristic cones.

The field equations of motion take the form of a conservation law, which is an expression of the symmetry of the action under shifts of the scalars,
\begin{equation}\label{eq:motion}
	\partial_\mu\left(\sqrt{-g}\,  \frac{dL_\alpha}{dX_\alpha} g^{\mu\nu} \partial_\nu \varphi_\alpha\right)=0.
\end{equation}
The energy-momentum tensor of the triad is simply the sum of the energy-momentum tensors of the individual scalars,
\begin{equation}\label{eq:EMT}
	T^\mu{}_\nu=\sum_\alpha\left(\frac{dL_\alpha}{dX_\alpha} \partial^\mu \varphi_\alpha \partial_\nu \varphi_\alpha -L_\alpha \,\delta^{\mu}{}_\nu\right).
\end{equation}

\section{Background}
We are interested in cosmological solutions that satisfy the cosmological principle, that is,  in  spatially homogeneous and isotropic universes,
\begin{equation}\label{eq:FRW}
	ds^2=a^2(\eta)\cdot (d\eta^2 - d\vec{x}^2).
\end{equation}
As opposed to what one would naively expect, homogeneous configurations do not require the field itself to be homogeneous. Since the action (\ref{eq:action}) only depends on the three scalars through their derivatives, the actual value of the fields is physically irrelevant, since only their gradients have  physical meaning. Therefore, homogeneous configurations only require that the field gradients be spatially constant. Hence, rather than assuming that the spatial gradients are zero, as is typically done, we shall study the ansatz
\begin{equation}\label{eq:ansatz}
	\varphi_\alpha= \Lambda^2 \, x^\alpha,
\end{equation}
which in conjunction with equation (\ref{eq:FRW}) results in
\begin{equation}
	X\equiv X_\alpha=-\frac{1}{2}\frac{\Lambda^4}{a^2}.
\end{equation}
Note that we are implicitly assuming that $\Lambda$ is a constant. Indeed, one can readily verify that for time-independent $\Lambda$,  the ansatz (\ref{eq:ansatz})  solves the equation of motion (\ref{eq:motion}). As a matter of fact, for the metric (\ref{eq:FRW}) the equation of motion reads 
\begin{equation}
\frac{dL_\alpha}{dX_\alpha}\cdot \left(\varphi_\alpha''+2\frac{a'}{a}\varphi_\alpha'-\Delta \varphi_\alpha\right)+\frac{d^2L_\alpha}{dX_\alpha^2}\cdot \left(X_\alpha'\varphi_\alpha'-\vec{\partial} X_\alpha \cdot \vec{\partial}\varphi_\alpha\right)=0,
\end{equation}
which is clearly satisfied by the ansatz (\ref{eq:ansatz}). Note that a prime denotes a derivative with respect to conformal time, and $\Delta\equiv \partial_i \partial^i $ is the flat space Laplacian.

Although the configuration (\ref{eq:ansatz}) does respect homogeneity, it does appear to violate isotropy, since the field gradients $\partial_i \varphi_\alpha =\Lambda^2\, \delta_{i \alpha}$ are not invariant under spatial rotations. However, this impression is misleading, because the contributions of each scalar to the energy momentum tensor (which is all gravity sees) may add to an isotropic energy momentum tensor. In fact, the total energy momentum tensor of the three scalars is given by
\begin{subequations}
\begin{eqnarray}
	T^0{}_0&=&-\sum_\alpha L_\alpha, \\
	T^0{}_i&=&0, \\
	T^i{}_j&=&\sum_\alpha 2 X_\alpha \frac{dL_\alpha}{dX_\alpha}\delta^i{}_\alpha \delta_{j \alpha}-L_\alpha \delta^i{}_j .
\end{eqnarray}
\end{subequations}
Therefore, if $dL_\alpha/d\log X_\alpha$ does not depend on $\alpha$, the energy-momentum tensor is invariant under spatial rotations. The simplest way to satisfy this condition is to assume that the Lagrangians are identical, 
\begin{equation}\label{eq:symmetry}
	L\equiv L_\alpha=L_\beta  \quad \quad (\alpha \neq \beta).
\end{equation}
In this case the theory has a $\mathbb{Z}_3$ symmetry that acts on the three scalar fields by permutation. For Lagrangians linear in $X$, this $\mathbb{Z}_3$ is just a subgroup of $O(3)$. 

If conditions (\ref{eq:symmetry}) are satisfied, the stresses are isotropic, and we can therefore identify
\begin{eqnarray}
	\rho&=&-3 L, \\
	p&=& 3L-2X\frac{dL}{dX}
\end{eqnarray}
as, respectively, the energy density and pressure of the scalar triad. In this sense, the three scalar fields are analogous to the group of three vectors described in \cite{Armendariz-Picon:2004pm}; the field gradients play then the role of of the spacelike vectors. Let us note that the ghost-free condition  $dL/dX>0$ implies, in conjunction with $X<0$,  that $\rho+p>0$. The scalar triad cannot behave like a phantom \cite{Caldwell:1999ew}.

For the sake of illustration, it is going to be convenient to have a class of concrete Lagrangians in mind. For our purposes it will suffice to consider a simple but sufficiently  general  set of models with Lagrangian
\begin{equation}\label{eq:L}
	L= C \cdot (D X)^n,
\end{equation}
where $C$ and $D$ are two arbitrary dimensionful constants and $n$ is an arbitrary exponent. In this case, the triad always behaves like a fluid with equation of state
\begin{equation}\label{eq:w}
	w=\frac{2n-3}{3}.
\end{equation}
The Lagrangian satisfies the stability conditions (\ref{eq:stability}) if $n>1/2.$

\section{Statistical Anisotropies} \label{sec:Anisotropies}
Our first goal is to  show that within the class of models that we are studying it is possible  to  create statistically anisotropic scalar primordial perturbations. We shall first define what we mean by ``scalar" and show that in our decomposition of the perturbations scalars, vectors and tensors generally couple to each other.  

\subsection{Perturbation Decomposition}

Let us start by considering metric perturbations in longitudinal gauge,
\begin{equation}\label{eq:metric perturbations}
	ds^2=a^2 \left[(1+2\Phi) d\eta^2+2 S_i \, d\eta\,  dx^i
	-(\delta_{ij}-2\Psi \, \delta_{ij}-h_{ij})dx^i dx^j\right],
\end{equation}
where $S_i$ is a transverse vector, and $h_{ij}$ is a symmetric transverse and traceless tensor. Essentially, we just have decomposed the metric perturbations in irreducible representations of a subgroup of symmetries of the FRW metric, namely, spatial rotations.  Rather than proceeding similarly with the triad perturbations, it will be instructive to expand\footnote{I thank Ignacio Navarro for pointing this out.}
\begin{equation}\label{eq:field perturbations}
	\delta\varphi_\alpha =\Lambda^2 \cdot (\partial_\alpha s+ v_\alpha),
\end{equation}
where $s$ is a scalar and $v_\alpha$ is a transverse vector, $\partial_\alpha v^\alpha\equiv \sum_\alpha \partial_\alpha v_\alpha=0$. Indeed, our scalar field background (\ref{eq:ansatz})  is not invariant under spatial rotations, but instead, remains unchanged under a $O(3)$ group of symmetries
\begin{equation}\label{eq:group}
	x_i\to R_{ij}\,  x^j, \quad \varphi_\alpha \to R^{-1}_{\alpha\beta}\, \varphi^\beta, \quad R\in O(3).
\end{equation}
The decomposition (\ref{eq:field perturbations}), as well as (\ref{eq:metric perturbations}), is a decomposition of perturbations in irreducible representations of the latter group of transformations. 

We shall not write down the different components of the linearized Einstein tensor, which can be found in many different monographs on the subject (e.g. \cite{Mukhanov:1990me}) .  The components of the linearized energy momentum tensor are
\begin{subequations}
\begin{eqnarray}
	\delta T^0{}_0&=&-2X\frac{dL}{dX} \left(3\Psi +\Delta s\right), \label{eq:00} \\
	\delta T^0{}_i&=& - 2 X \frac{dL}{dX} \left(S_i+\partial_i s'+v'_i\right), \label{eq:0i} \\
	\delta T^i{}_j&=&2X \frac{dL}{dX}\cdot\left[\Psi \delta_{ij}+(2\partial_i \partial_j-\delta_{ij}\Delta) s+\partial_i v_j +\partial_i v_j+h_{ij}\right]+ \nonumber \\
	& &{}+2X^2\frac{d^2L}{dX^2}\sum_\alpha \delta_{\alpha i}\delta_{\alpha j}
	\left[2\psi+2\partial_\alpha^2 s+2\partial_\alpha v_\alpha+h_{\alpha\alpha}\right]
	\label{eq:ij}.
\end{eqnarray}
\end{subequations}
Although equations (\ref{eq:00}) and (\ref{eq:0i}) appear to suggest that scalar, vector and tensor perturbations decouple, closer inspection of equation (\ref{eq:ij}) shows that this is not the case whenever $d^2L/dX^2 \neq 0$. Consider for instance the decomposition in irreducible representations  of the last term in equation (\ref{eq:ij}),
\begin{equation}
	 \sum_\alpha \delta_{\alpha i}\delta_{\alpha j} \, h_{\alpha \alpha }\equiv A \delta_{ij}+\partial_i \partial_j B+\partial_i C_j+\partial_j C_i+D_{ij}.
\end{equation}
The scalar $A$ is given by $A=-\frac{1}{\Delta} \sum_\alpha \partial_\alpha^2\,  h_{\alpha\alpha}$  \cite{Arnowitt:1962hi}, 
which does not vanish and does depend on the tensor $h_{ij}$. Hence, the linearized Einstein equations couple scalars to vectors and tensors. The same conclusion holds if the triad perturbations are expressed in terms of the variables $\delta\varphi_\alpha$. The violation of the decomposition theorem \cite{Kodama:1985bj} was noted in a similar context in \cite{Armendariz-Picon:2004pm}. Note by the way that the term $2X dL/dX h_{ij}$ in equation (\ref{eq:ij}) gives a mass to the graviton \cite{Dubovsky:2004sg}.

There are two cases where scalars, vectors and tensors can be  decoupled, namely, when $X^2 d^2L/dX^2=0$ or when the metric perturbations themselves are negligible. In the first case, the group of transformations (\ref{eq:group}) is not just a symmetry of the background, but also a symmetry of the action. As a result of this enhanced symmetry the triad perturbations turn out to be  statistically isotropic, as we shall see below. We are hence led to the limit  in which the metric perturbations are much bigger than the triad perturbations, and are thus negligible.   We expect this to be the case, for example,  if the expansion of the universe is sufficiently close to de Sitter.

\subsection{Primordial Perturbations}
\label{sec:Primordial Perturbations}
We shall study now the perturbations in the triad created during a stage of seeding in the early universe.  At this point it is not important to specify what  is the dominant component of the universe during that stage; it could be the triad itself, or it could be a different scalar, like in curvaton models \cite{Linde:1996gt}.  We shall simply assume that the early universe inflates, and, for definiteness, that its effective equation of state $w$ is constant,
\begin{equation}
	a=a_0 \left(\frac{\eta}{\eta_0}\right)^p, \quad \text{where} \quad p=\frac{2}{1+3w}.
\end{equation}

In order to calculate the primordial spectrum of metric perturbations, we have to specify some of the details of the transition between the seeding stage and a radiation dominated universe. If reheating is instantaneous and occurs along an hypersurface of constant $\sum_\alpha X_\alpha$, we find that after a stage of inflation with equation of state $w$ the universe contains metric perturbations (in longitudinal gauge)  even though they were negligible during inflation \cite{Deruelle:1995kd},
\begin{equation}\label{eq:delta}
	\Phi\approx \frac{3w-1}{18}\,\delta, \quad \text{where} \quad 
	\delta\equiv \sum_\alpha \frac{1}{2}\frac{\delta X_\alpha}{X}=\frac{1}{\Lambda^2}\sum_\alpha \partial_\alpha \delta\varphi_\alpha.
\end{equation}
Similarly, if the decay rate of the component that drives the seeding stage depends on the squared gradients of the triad $X_\alpha$, perturbations in the latter  will be converted into adiabatic primordial perturbations at the end of the seeding stage \cite{Dvali:2003em}.  In our theory, by assumption, the Lagrangian is symmetric under shifts of the scalars and permutations, so the decay rate has to be of the form $\Gamma_\mathrm{tot}=\sum_\alpha \Gamma(X_\alpha)$. As a consequence, we expect primordial perturbations of the order
\begin{equation}
		\Phi\approx \frac{1}{9}\frac{\delta \Gamma_\mathrm{tot}}{\Gamma_\mathrm{tot}}=\frac{2}{27} \frac{d\log \Gamma}{d\log X} \, \delta,
\end{equation}
where $\delta$ has been defined in equation (\ref{eq:delta}). The term proportional to $d \log \Gamma/d\log X$ is typically of order one, an hence, primordial perturbations are again proportional to $\delta$.  In the following we shall calculate the spectrum of $\delta$.

\subsection{Power Spectra}
We are ready now to calculate the action for the cosmological perturbations. As mentioned above, we are neglecting metric perturbations, so all we have to do is expand the action of the scalars to quadratic order in the presence of an unperturbed FRW  metric. Proceeding in this manner we arrive at
\begin{equation}\label{eq:delta S}
	\delta S=\sum_\alpha \int d^4 x \, \frac{a^2}{2} \left[\frac{dL}{dX}\left(\delta\varphi_\alpha'{}^2-\partial_i \delta\varphi_\alpha \partial^i \delta\varphi_\alpha\right)-2X \frac{d^2L}{dX^2} \partial_\alpha\delta\varphi_\alpha \partial_\alpha \delta\varphi_\alpha\right],
\end{equation}
from which one can directly derive the stability conditions (\ref{eq:stability}). Note that we have expressed the field perturbations in terms of the perturbed fields $\delta\varphi_\alpha$. These are the fields that decouple in the absence of metric perturbations. When we substitute equation (\ref{eq:field perturbations}) into the action (\ref{eq:delta S}), we find that the term proportional to $d^2L/dX^2$ couples the vector $v_\alpha$ to the scalar $s$.  The very same term manifestly breaks rotational invariance in the action, and will be ultimately responsible for the generation of statistically anisotropic perturbations.

In order to quantize the perturbations, it is convenient to further decompose them in Fourier modes of an appropriately rescaled scalar field (with periodic boundary conditions),
\begin{equation}\label{eq:v}
	v_\alpha\equiv b \cdot \delta\varphi_\alpha, \quad \text{where} \quad b^2=a^2\cdot \frac{dL}{dX}.
\end{equation}
Note that $b$ plays the role of an ``effective" scale factor, sometimes referred to  as the pump field.  Substituting the expansion (\ref{eq:v}) into the action (\ref{eq:delta S}), and employing the particular class of models in (\ref{eq:L}) we arrive at
\begin{equation}\label{eq:v a}
	 \delta S_\alpha =V \sum_{\vec{k}} \int d\eta\, \frac{1}{2}\left[v_k' v_{-k}' -\left(\omega_\alpha^2-\frac{b''}{b}\right)v_{k} v_{-k}\right],
\end{equation}
where
\begin{equation}\label{eq:dispersion}
	\omega_\alpha^2=\vec{k}^2+2(n-1) k_\alpha^2
\end{equation}
is the anisotropic dispersion relation of the field perturbation $\delta\varphi_\alpha$ and $V$ is the comoving volume of the universe..   As the reader may have expected, the dispersion relation is isotropic only if $d^2 L / dX^2$ vanishes ($n=1$). 

The quantization of (\ref{eq:v a}) proceeds along the conventional lines \cite{Birrell:1982ix}. The properly normalized mode functions are
\begin{equation}
	v_k^\alpha=\sqrt{\frac{-\pi \eta}{4V}} H_\nu^{(1)}(-\omega_\alpha \eta), 
\end{equation}
where $H_\nu$ is the Bessel function of the first kind and
\begin{equation}\label{eq:nu}
	\nu=\frac{1-2(2-n)p}{2}.
\end{equation}

Using the definition of $\delta$ in equation (\ref{eq:delta}), and the relation between $v_\alpha$ and $\delta\varphi_\alpha$ in equation (\ref{eq:v}) it is now straightforward to calculate the spectrum of $\delta.$ For $\nu>0$ the spectrum  is constant on large scales $(\omega_\alpha \eta \ll 1)$. Therefore, in that limit it can be evaluated at any convenient time, which we choose to be the time at which a reference mode $k_*$ crosses the horizon, $k_*/a_*=H_*$. On large scales the primordial spectrum is then 
\begin{equation}\label{eq:spectrum}
	\mathcal{P}_\delta = \frac{\Gamma^2(\nu)}{4\pi^3} \cdot \left(-1-3w\right)^{2\nu-1} \cdot\frac{H_*^4}{\rho_*+p_*} \left(\frac{k}{k_*}\right)^{5-2\nu} 
	\sum_\alpha\, \left(\frac{k_\alpha}{k}\right)^2 \left(\frac{\omega_\alpha}{k}\right)^{-2\nu},
\end{equation}
where the dispersion relation $\omega_\alpha$ is given by equation (\ref{eq:dispersion}), and the index $\nu$ is given by equation (\ref{eq:nu}). The equation of state of the dominant component during inflation is $w$, and $\rho_*$ and $p_*$ denote, respectively, the energy density and pressure of  the triad at the time of crossing.  

If $n\neq 1$, primordial perturbations are statistically anisotropic, even though the background energy-momentum tensor and the spacetime metric are invariant under rotations. In any case, the primordial spectrum still possesses certain remnant symmetries. In particular,  it is scale free and  invariant under spatial inversions, $\vec{k}\to -\vec{k}$. The scaling symmetry follows from the structure of the power spectrum, which is a product of two functions, $\mathcal{P}(\vec{k})=\mathcal{P}(k) \cdot \Omega (\hat{k})$, where $\hat{k}=\vec{k}/k$.  One could ask whether this factorization is generic, or whether it  is just particular feature of our set of models. By tracing back the steps we took to derive the power spectrum, we see that the origin of this property is the form of the dispersion relation $\omega_\alpha$. If  $\omega_\alpha$ consists of terms with different powers of $k_\alpha$, $\mathcal{P}_\delta(\vec{k})$ cannot be factorized.  Similarly, it is easy to see that the parity symmetry arises from the symmetry of the dispersion relation. Dispersion relations with odd powers of $\vec{k}$ would lead to parity violations.

As we shall argue below, the factorization of the primordial spectrum allows us to identify an effective spectral index of the perturbations, $n_s-1=5-2\nu$.  In the following we shall focus on the phenomenologically preferred case of  scale invariant perturbations, $n_s=1$. The spectrum is scale invariant in this generalized sense if $\nu=5/2$, which implies
\begin{equation}
	w=\frac{n-3}{3}.
\end{equation}
We can compare this condition with the equation that relates the expansion of the universe to the parameter $n$ when the triad dominates the energy density of the universe, equation (\ref{eq:w}). Only for $n=0$ are both mutually consistent. However, for this value of the parameter  perturbations are quantum-mechanically unstable. Hence, in this class of models, the component that drives inflation cannot be the triad.  Because stability requires $n>1/2$, and since in our context the seeding requires a stage of inflation, $-1<w<-1/3$, the range of values of exponents compatible with a scale invariant spectrum is  $1/2<n<2$.  These conclusions are closely tied to our class of Lagrangians (\ref{eq:L}), and could be relaxed in other types of models. 

\subsection{Observational Signatures}
Let us  briefly describe what type of signatures a spectrum of the form (\ref{eq:spectrum}) would leave on the cosmic microwave background.  The impact  of an arbitrary statistically anisotropic spectrum of primordial perturbations on the CMB was calculated in \cite{Armendariz-Picon:2005jh}, so all we have to do is apply the formalism developed in the last reference, which the reader may  consult for further details.

Our starting point is the decomposition of the primordial spectrum in spherical harmonics,
\begin{equation}\label{eq:decomposition}
	\mathcal{P}_\Phi=\sqrt{4\pi} \sum_{\ell m} \mathcal{P}_{\ell m}(k) Y_{\ell m}(\hat{k}).
\end{equation}
The functions $\mathcal{P}_{\ell m}$ quantify the power in the harmonic $\ell m$ as a function of scale $k$. For statistically isotropic perturbations, only the scalar component $\mathcal{P}_{00}$ differs from zero. Certainly, this is not the case in our class of models.  In particular, comparing equation (\ref{eq:spectrum})  with equation (\ref{eq:decomposition}), and remembering that the spectra of $\delta$ and $\Phi$ just differ by an overall constant,  we arrive at
\begin{equation}
	\mathcal{P}_{\ell m}= \frac{A^2}{\sqrt{4\pi}} \cdot  \left(\frac{k}{k_*}\right)^{5-2\nu}  
	\int d^2\hat{k}\,  Y^*_{lm} (\hat{k}) \cdot \sum_\alpha  \frac{\hat{k}_\alpha^2}{\left(1+ 2(n-1)\hat{k}_\alpha^2\right)^\nu},
\end{equation}
where $A$ is an overall amplitude we shall not worry about.  Because  all the harmonics share the same power of $k$ we can identify ${n_s=6-2\nu}$ as the spectral index of the primordial spectrum. Since the spectrum is invariant under spatial inversions, harmonics with an odd angular momentum quantum number vanish, $\mathcal{P}_{\ell m}=0$ for odd $\ell$. Similarly, because $\Omega(\vec{k})$ is invariant under reflections on the $k_x-k_y$ plane, $\mathcal{P}_{\ell m}=0$ for odd $m.$
\begin{figure}
  \begin{center}
 	\includegraphics[scale=0.35]{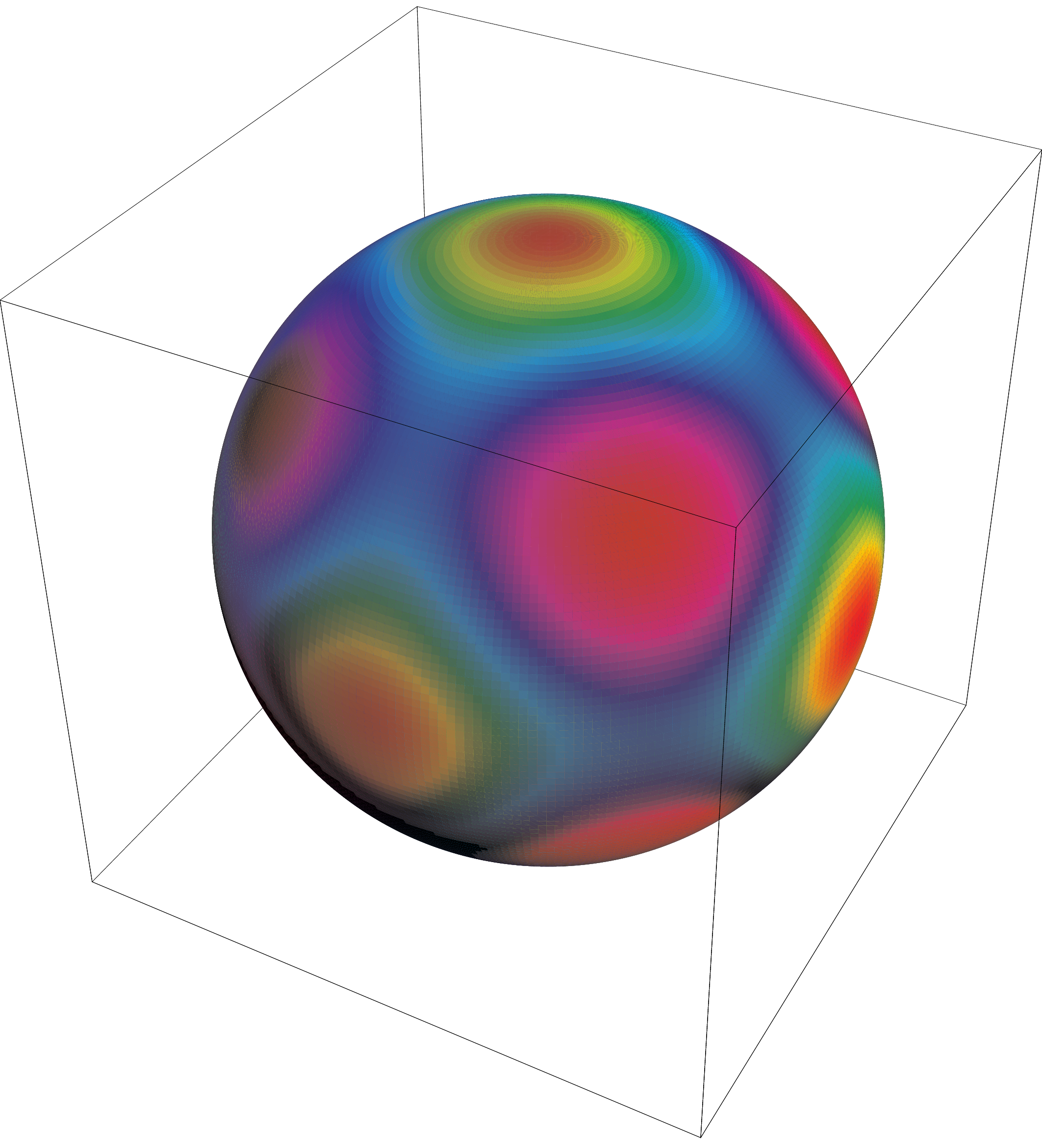}
  \end{center}
  \caption{A density plot of the function $\mathcal{P}_\Phi(\hat{k})$ for $n=3/2$. }
   \label{fig:omega}
\end{figure}

The different harmonics of the power spectrum are connected to the statistical properties of the CMB. In particular, the expectation value of the estimator typically used to measure the angular power spectrum\footnote{In practice, sky cuts and noise in the temperature maps require the use of somewhat different estimators \cite{Hinshaw:2006ia}.} is
\begin{equation}
	C_\ell\equiv \frac{1}{2\ell+1} \sum_m \langle a^*_{\ell m} a_{\ell m} \rangle =\int \frac{dk}{k} \Delta^2_\ell (k) \, \mathcal{P}_{00}(k),
\end{equation}
where $\Delta_\ell$ is the conventional transfer function \cite{Ma:1995ey}. Therefore, the angular power spectrum, as defined above,  does not contain any information about statistical anisotropies, since it is only sensitive to the scalar component of the power spectrum $\mathcal{P}_{00}$. If the spectrum is scale invariant (in our generalized sense), statistically anisotropic perturbations will fit the angular power spectrum as well as  statistically isotropic ones.

The anisotropic components of the power spectrum affect the non-diagonal  two-point functions, 
\begin{equation}
	\langle a^*_{\ell_1 m_1} a_{\ell_2 m_2}\rangle = (-i)^{\ell_2-\ell_1} \sum_{\ell m} D(\ell_1, m_1; \ell, m | \ell_2, m_2)
	\int \frac{dk}{k} \Delta_{\ell_1}(k) \Delta_{\ell_2}(k) \mathcal{P}_{\ell m}(k),
\end{equation}
where $D$ is a product of Clebsch-Gordan coefficients \cite{Armendariz-Picon:2005jh}. Hence, statistically anisotropic perturbations yield correlations between the different spherical coefficients of the CMB. A color-coded contour plot of the power spectrum $\mathcal{P}_\Phi$ on the unit sphere ($\vec{k}\in S^2$) is shown in figure \ref{fig:omega}.  From the figure, it is apparent that the smallest structures in $\mathcal{P}_\Phi(\hat{k})$ have angular sizes of about $\pi/6$. Hence, we expect its multipole components to be insignificant beyond $\ell\approx 8$.  In table \ref{tab:multipoles} we also list the first few non-vanishing multipoles of our fiducial model with $n=3/2$. Note that there is no dipole, which is the multipole we would expect from just one preferred spatial direction.  Instead, the dominant multipole is $\ell=4$.
\begin{center}
\begin{table}
 $
 \begin{array}[t]{|c|c|c|}
\hline
\ell & m & \mathcal{P}_{\ell m} \\
 \hline
 4&  0 & -0.20  \\
 & 4 & -0.12  \\
 \hline
 \end{array}
 $
 $
\begin{array}[t]{|c|c|c|}
\hline
\ell & m & \mathcal{P}_{\ell m} \\
 \hline
 6&  0 & 0.01 \\
 & 4 & -0.03 \\
 \hline
 \end{array}
 $
 $
 \begin{array}[t]{|c|c|c|}
\hline
\ell & m & \mathcal{P}_{\ell m} \\
 \hline
8 &  0 & 0.02 \\
 & 4 & -0.006 \\
 & 8 & -0.009 \\
 \hline
 \end{array}
 $
 $
 \begin{array}[t]{|c|c|c|}
\hline
\ell & m & \mathcal{P}_{\ell m} \\
 \hline
10 &  0 & 0.001 \\
 & 4 & -0.001 \\
 & 8 & -0.002 \\
 \hline
 \end{array}
$
 \caption{The first  non-vanishing multipole components of the power spectrum  for $n=3/2$.}
 \label{tab:multipoles}
\end{table}
\end{center}
Statistically anisotropic primordial perturbations of this type could a  priori explain the anomalous alignment of the quadrupole and the octopole of the CMB \cite{deOliveira-Costa:2003pu}. Although no evidence for statistical anisotropy in the CMB has been found in the model-independent analysis of  \cite{Hajian:2006ud}, this does not imply  that statistically anisotropic primordial perturbations are ruled out. Bounds on the amplitude of the different power spectrum components $\mathcal{P}_{\ell m}$ do not exist (as far as the author knows), so one cannot exclude this type of perturbations without further analysis.  The situation here resembles to some extent the status of non-adiabatic perturbations. Although there is no evidence for non-adiabatic primordial perturbations, current bounds still allow significant isocurvature components \cite{Bean:2006qz}. 

\section{Statistical Inhomogeneity}
In the light of these results, the reader may wonder whether it is also possible to seed statistically inhomogeneous primordial perturbations in a Friedman-Robertson-Walker universe. Indeed, it is possible to do so by dropping the requirement that the triad be invariant under shifts. Consider for instance the following matter Lagrangian,
\begin{equation}
	\mathcal{L}_m=\frac{1}{2}\partial_\mu \chi \partial^\mu\chi
	-\frac{1}{2}\sum_{a=1}^3 m^2(\varphi_\alpha)\chi^2+L_\psi(g_{\mu\nu},\psi, \chi).
\end{equation}
It contains an additional scalar $\chi$, whose field-dependent mass breaks the shift symmetry of the theory.  The $\mathbb{Z}_3$  symmetry under permutations of the triad is still preserved, and this time we assume   that only the scalar field $\chi$ has couplings to the standard model matter fields $\psi$. Because the shift symmetry is broken, radiative corrections should lead to many additional field-dependent terms in the Lagrangian. Their absence is likely to require certain degree of fine-tuning. 

We shall assume that the field $\chi$ is at the bottom of its potential, $\chi\equiv 0$. This is not only consistent, but also expected if the mass of the field  is sufficiently big. Then, the scalar $\chi$ does not contribute to the energy density of the universe, and does not affect the equations of motion of the triad. In particular, the configuration (\ref{eq:ansatz}) is still a cosmological solution, and equation (\ref{eq:EMT}) still gives its energy-momentum.

\subsection{Power Spectra}
Let us study now what type of primordial perturbations could be produced under such circumstances. First, note that the field $\chi$ couples to matter, so perturbations in the former can be converted into adiabatic primordial perturbations at the end of a seeding stage by the Dvali-Gruzinov-Zaldarriaga  mechanism \cite{Dvali:2003em} we discussed in Section \ref{sec:Primordial Perturbations}. Hence, it will suffice to calculate the power spectrum of $\chi$. As before, we shall neglect metric perturbations.

In order to obtain the action for the perturbations of $\chi$, let us work again in the Fourier space of the rescaled variable $v=a \chi$. In terms of this variable, the action for the scalar $\chi$ is
\begin{equation}\label{eq:}
	S_\chi =\int  d\eta\, \sum_{\vec{k},\vec{l}}  \frac{V}{2}\left[\delta_{k l} \left( v_k' v_{-l}' - \vec{k}^2\,   v_{k}\,  v_{-l}+\frac{a''}{a} v_{k} \, v_{-l}\right)+a^2 \, m^2_{l-k} \, v_{k}\, v_{-l}\right],
\end{equation}
where $m^2_k$ are the Fourier modes of the squared mass of $\chi$ evaluated  along the configuration of the triad fields,
\begin{equation}
	m^2_k=\frac{1}{V}\int d^3x \, e^{-i \vec{k}\vec{x}} \sum_\alpha m^2(\Lambda^2 \, x^\alpha),
\end{equation}
which do not depend on time. The form of the action manifestly shows that  the Fourier components of the field $\chi$ do not decouple.  Consider for instance the equation of motion of the mode $\vec{k}$,
\begin{equation}
	v_k''+\left(\vec{k}{}^2-\frac{a''}{a}\right) v_k+\sum _{l} a^2\,  m^2_{k-l} \, v_{l}=0.
\end{equation}
It only contains that mode if $m^2_{k-l}\propto \delta_{kl}$, that is, if $m^2$ is spatially constant.  But in that case the action is invariant under spatial translations, so it is not possible to seed statistically inhomogeneous perturbations.   We therefore encounter again  a link between violations of the decomposition principle and the generation of statistically asymmetric perturbations. 

Because different Fourier modes are not decoupled, the quantization of this theory requires more effort. Instead of diagonalizing the action, we shall treat the mass as a perturbation. This is a good approximation for a sufficiently light field, $m^2/H^2\ll 1$, and a sufficiently small  number of e-folds of inflation \cite{Weinberg:2006ac}. To first order in perturbation theory, a non-zero mass term generates a correction to the zeroth order power spectrum \cite{Weinberg:2005vy}
\begin{equation}\label{eq:correction}
	\Delta \langle v^\dag_k(\eta) v_{l}(\eta)\rangle =
	 i \int\limits_{-\infty}^{\eta} d\tilde{\eta} \, 
	 \langle[H_\mathrm{int}(\tilde{\eta}), v^\dag_k(\eta) v _{l}(\eta)]\rangle,
\end{equation}
where $H_\mathrm{int}$ is the interaction Hamiltonian,
\begin{equation}
	H_\mathrm{int}=\frac{1}{2}\sum_{kl} a^2 \, m^2_{l-k} \, v_{k}\, v_{-l}.
\end{equation}
Note that we are working in the interaction picture, where the operators $v_k$ carry the free (massless) time evolution.

For simplicity, we shall assume a de Sitter expansion of the universe, $p=-1$, in which the properly normalized mode functions are 
\begin{equation}
	v_k=\frac{e^{-i k \eta}}{\sqrt{2k V}}\left(1-\frac{i}{k\eta}\right).
\end{equation}
Then, at late times, the integral in equation (\ref{eq:correction}) diverges with the logarithm of the scale factor \cite{Weinberg:2006ac}.   We shall restrict our calculation to the leading logarithmic correction to the correlation function in the limit of late times. Isolating that contribution,  after a fair amount of calculation with exponential integrals, we find that the two-point function of $\chi$ is given by
\begin{equation}
	 \frac{V \cdot k^3}{2\pi^2} \langle \chi^\dag _k\,  \chi_{l}\rangle
	 \approx\frac{H^2}{4\pi^2} \delta_{k l} 
	 + \frac{k^3}{12\pi^2} \cdot \left(\frac{1}{k^3}+\frac{1}{l^3}\right) 
	 m^2_{l-k}  \, \log \frac{k+l}{a \, H}.
\end{equation}
Hence, the two-point function is not proportional to a momentum-conserving Kronecker delta,  a momentum dependence that signals the breakdown of statistical homogeneity \cite{Armendariz-Picon:2005jh}. In this case, the conventional notion of  ``power spectrum" $\mathcal{P}_\chi(\vec{k})$  is not defined. 

\section{Summary and Conclusions}
In this article we have explored the relation between the symmetries of a Friedman-Robertson-Walker universe and the statistical symmetries of its perturbations. Our starting point has been the cosmological principle, the homogeneity and isotropy of the universe, which we have taken for granted. We have then explicitly shown that it is possible to generate statistically anisotropic and inhomogeneous primordial perturbations if the early universe  contains a  scalar triad.  The triad is a set of three scalar fields with non-vanishing spatial gradients. Its properties are essentially determined by the cosmological symmetries, which require that the gradients be mutually orthogonal and spatially constant. Although this configuration is not isotropic, it turns out that its energy momentum tensor  is invariant under rotations, and thus, is compatible with the symmetries of a homogeneous and isotropic FRW universe. 

In our seeding scenario, statistically anisotropic or inhomogeneous primordial perturbations arise from the breaking of rotational or translational invariance by the triad configuration, which also prevents the decoupling of the different modes of cosmological perturbations.  Because rotational and translational invariance must be ultimately broken in any model that singles out  particular directions or locations,  it is likely that the connection between statistically  anisotropic or inhomogeneous perturbations  on one hand and violations of the decomposition theorem on the other is generic. 

We have discussed a particular class of simple models that lead to a statistically anisotropic primordial spectrum of perturbations.  Even though in general it is not possible to do so, in our example we can define a single spectral index that uniquely captures the scale dependence of the primordial power. Requiring scale invariance of the created perturbations then imposes additional constraints on the scalar field Lagrangian, and implies---at least for the restricted class of Lagrangians that we have studied---that the scalar triad cannot be the dominant component during the seeding stage in the early universe.  The resulting  angular dependence of the primordial spectrum is quite rich, and does not simply correspond to the dipole pattern one would associate to a single preferred direction.

It is also possible to generate statistically inhomogeneous perturbations during a stage of inflation. In this context, the breaking of translation invariance arises from the spatial variations of the triad, and the explicit breaking of a shift symmetry by the coupling of the triad to matter. Because the definition of power spectrum implicitly assumes statistical homogeneity, we cannot  quote a power spectrum in the conventional sense, since the two-point function now depends on two momentum variables.   In the class of models we have considered, the kinetic terms of the triad did not play any role, and could be taken to be canonical. 

Whether or not primordial perturbations are statistically homogeneous and isotropic is a question that should be settled by experiment, not by theoretical prejudice.  Certain anomalies of the CMB temperature anisotropies indeed suggest for instance that cosmological perturbations are statistically anisotropic. Although these anomalies could have a primordial origin, we have not explored this possibility here. Instead,  we have simply argued that statistically anisotropic or inhomogeneous perturbations do not necessarily clash with our understanding of the origin of structure in the universe. 

\begin{acknowledgments}
It is a pleasure to thank Sergei Dubovsky and Ignacio Navarro for useful conversations. This work is supported in part by the National Science Foundation under grant PHY-0604760. 
\end{acknowledgments}

\end{document}